\documentclass[aps,prc,showpacs,twocolumn]{revtex4}
\usepackage{graphicx}
\usepackage{dcolumn} 
\usepackage{bm}
\usepackage{hyperref}
\usepackage{natbib}
\usepackage{amsmath}
\usepackage{multirow}
\begin{document}
\title{Effect of Pauli principle on the deformed QRPA calculations and its consequence in the $\beta$-decay calculations of deformed even-even nuclei}
\author{Dong-Liang Fang}
\affiliation{College of Physics, Jilin University, Changchun, Jilin 130012, China}
\begin{abstract}
In this work, we take into consideration of Pauli Exclusion Principle(PEP) in the quasi-particle random phase approximation (QRPA) calculations for the deformed systems by replacing the traditional Quasi-Boson Approximation(QBA) with the renormalized one. With this new formalism, the parametrization of QRPA calculations has been changed and the collapse of QRPA solutions could be avoid for realistic $g_{pp}$ values. We further find that the necessity of renormalization parameter of particle-particle residual interaction $g_{pp}$ in QRPA calculations is due to the exclusion of PEP. So with the inclusion of PEP, we could easily extend the deformed QRPA calculations to the less explored region where lack of experimental data prevent effective parametrization of $g_{pp}$ for QRPA methods. With this theoretical improvement, we give predictions of weak decay rates for even-even isotopes in the rare earth region and the results are then compared with existing calculations. 
\end{abstract}
\maketitle
\section{introduction}
Rapid neutron capture process (r-process) plays an important role on the formation of our solar system element abundance and it is thought to be responsible for productions of most heavy elements\cite{BBF57} beyond $^{56}$Fe. For, interest has been attracted  on the determination of the site for this process, various models have been proposed, {\it i.e.} the hot high entropy neutrino wind model\cite{AJS06}, the cold neutron star mergers model\cite{FRT99}, {\it etc}. This process requires large neutron-to-seed-nuclei ratios at the initial phase, while the r-process path and the final abundance are determined by the competitions between the neutron capture rates and the weak decay rates at the late stage. In this sense, precise values of weak decay rates especially for isotopes along or around the r-process path become crucial for the simulation of this process. Unfortunately most of these isotopes are neutron-rich and very unstable, this makes most of them out of the reach of current rare-isotope experiments. Thus, one needs to rely on accurate theoretical predictions of these properties in order to understand what is happening during these processes. Efforts have been put over this topic over decades, {\it i.e.} the global calculations microscopic or gross\cite{MR89,MNK97,MPK03}. 

Besides these global calculations over the whole nuclide chart, there are also numerous approaches devoted to calculating nuclei in specific regions, {\it i.e.} the Large Scale Shell Model(LSSM) calculations have been adopted for calculations for magic or semi-magic nuclei\cite{SYK11,ZCC13} which are usually with spherical shapes so that spherical symmetry could be imposed. For these nuclei, methods such as spherical QRPA with various kinds\cite{Bor03,NNL12,MVR07} are also applicable, and their agreement among each other has been improved steadily. Not all the nuclei are in spherical shapes, for open shell nuclei lies deep in the center of the square area surrounded by the magic lines on the nuclide chart, permanent deformations have been observed and for these nuclei, spherical symmetry are heavily broken. Simulation shows that the r-process path goes through two such deformed regions, the first is the Kr-Mo region (region I hereafter,) and another is the neutron-rich rare earth region(region II). For region I, exotic neutron-rich isotopes far from stability line have been measured recent years({\it e.g.} for Zr isotopes, those with 16 neutron richer than stable ones have been measured ), see for example\cite{Nis11,Lor15}. This ease the r-process simulation and could improve our understandings of the shape of A$\sim$140 peak of the abundance pattern. Efforts from theoreticians have also been put on calculating rates of these nuclei with , for example, the deformed QRPA methods either self-consistently with the Skyrme \cite{SP10,Yos13} or Gogny\cite{MPG14} forces or the traditional non-self-consistent case with realistic interactions\cite{FBS13,NR14}. On the other hand, situation for region II is less satisfying, for these deformed isotopes, due to their heavy mass, difficulty of their production and storage poses obstacles from measuring their basic properties such as mass or decay rates. Currently only a small number of them have been measured, for example, the heavily deformed Nd isotopes of six neutrons beyond quasi-stable $^{150}$Nd nucleus have been experimentally accessed. 

As pointed out by\cite{MMS12,MMS11}, weak decay properties of these nuclei could be crucial for a long-bothering problem of elements abundance, the rare earth peak. If one observes the abundance pattern, besides the two giant peaks near A$\sim$140 and A$\sim$190 caused by two shell gaps around neutron number N$=82,126$, we could find also a bump (or peak) near A$\sim$165. The formation of this peak has been investigated in\cite{MMS12,MMS11} with different mechanisms such as quenched shell gaps or fission back from heavy nuclei. They show that to distinguish from these different mechanisms, we need more accurate nuclear data such as masses or weak decay properties for hundreds of nuclei in the deformed rare earth region. As the absence of experimental data mentioned above, one needs to resort to the theories. Not too many calculations have been done for this region compared to region I. There are global calculations from \cite{MPK03}, but no alternatives are available. \cite{MTZE14} have provided some of the rates with fitted parameters from well measured spherical nuclei. In this work, we introduce the renormalized Quasi-particle Random Phase Approximation(rQRPA) methods for the calculations of even-even nuclei by considering the Pauli Exclusion Principle(PEP) approximately. We find out that the deviation of a renormalization parameter of the realistic particle-particle residual interaction (or the namely the strength of the interaction) $g_{pp}$ from 1 is largely due to the ignorance of the PEP. With our approximate treatment of PEP in QRPA approaches by the boson mapping technique, one could use bare G-matrix elements for the realistic nuclear structure calculations, that is $g_{pp}=1$. We compare our results to the measurements as well as calculations from \cite{MPK03}. Decent agreements have been achieved in both regions discussed above. This article is arranged as follows, in section II we give detailed formalisms of the method, then we discuss the parameters in section III, followed by results and discussions in Section IV, and finally the conclusion.

\section{formalism}
For the deformed system, single particle wave functions are generally described in the intrinsic frame, they are usually expanded on specific basis\cite{DPP69}: 
\begin{eqnarray}
|\Omega_\tau\rangle=\sum_{\{N\}\Lambda \Sigma} b^{\Omega_\tau}_{\{N\}} |\{N\}\Lambda\rangle |\Sigma\rangle 
\end{eqnarray}
Here $\{N\}$ is a set of quanta, {\it i.e.} with the deformed Harmonic Oscillator (H.O.) basis, theres are $\{n_z,n_\rho\}$. Here, $\Lambda$ is projection of orbital angular momentum on symmetry axis $z$ of the axially deformed system, and $\Sigma$ are the projection of spin on this axis.

The QRPA method is based on quasi-particle representation which are derived by solving either the BCS or HFB equations. In our calculations, we use BCS method to treat the nuclear pairing, under this scenario, the quasi-particle operators can be expressed as:
\begin{eqnarray}
\alpha_\tau=u_\tau c^\dagger_\tau+v_\tau \tilde{c}_\tau
\end{eqnarray}
Here u's and v's are the so-called BCS coefficients.

One then defines the proton-neutron QRPA(pn-QRPA) phonon for the charge-exchanging case as:
\begin{eqnarray}
Q_{K^\pi}^{m\dagger}=\sum_{pn} X^{m}_{pn} A^{\dagger}_{pn}-Y^{m}_{pn} \tilde{A}_{pn} \\
\Omega_{p}+\Omega_n=K \nonumber
\end{eqnarray}
In order to get these forward and backward amplitude X's and Y's, one then uses the variational methods as explained in \cite{RS80,HS67} to derive the so-called pn-QRPA equations\cite{SPF03,YRF08}. 
%added part
Here the two quasi-particle creation operator is defined as: $A_{pn}=\alpha_p^\dagger\alpha_n^\dagger$.

The commutation relations $[A_{pn},A^\dagger_{p'n'}]=\delta_{pp'}\delta_{nn'}$ are usually used in the derivation of these QRPA equations, this is the so-called Quasi-Boson Approximation (QBA) which treats the combination of two quasi-particle creation operators as a boson operator. However, this commutation relation is not exact since it neglects the Pauli exclusion principle (PEP) for multi-fermion system. In this work, we go beyond this approximation by replacing this commutation relation with the one used previously in\cite{CDS94,TS95,SSF96} for spherical systems:
\begin{eqnarray}
[A_{pn},A^{\dagger}_{p'n'}]&=&\delta_{pp'}\delta_{nn'}-\delta_{pp'}\langle 0^+_{QRPA}|\alpha^\dagger_{n'}\alpha_n|0^+_{QRPA}\rangle \nonumber \\
&&-\delta_{nn'}\langle 0^+_{QRPA}|\alpha^\dagger_{p'}\alpha_p|0^+_{QRPA}\rangle \nonumber \\
&=& \mathcal{D}_{pp',nn'}
\end{eqnarray}
This is the so-called renormalized quasi-boson approximation(rQBA). A general assumption here is that under QRPA vacuum, only diagonal terms exist, this implies that only nucleon pairs with $K^\pi=0^+$ contributes to the coefficients $\mathcal{D}$. By applying this new commutation relation, we could obtain the QRPA equations in the form:
\begin{eqnarray}
\left(\begin{array}{cc}
\mathcal{A}&\mathcal{B}\\-\mathcal{B}^{*}&-\mathcal{A}^{*}
\end{array}\right)
\left(\begin{array}{c}
X\\Y
\end{array}\right)=\omega \mathcal{D}
\left(\begin{array}{c}
X\\Y
\end{array}\right)
\end{eqnarray}
Here the QRPA matrices are defined as in \cite{SPF03,YRF08} $\mathcal{A}_{pn,p'n'}=[A_{pn},[H,A^\dagger_{p'n'}]]$ and $\mathcal{B}_{pn,p'n'}=[\tilde{A}_{pn},[H,\tilde{A}_{p'n'}]]$. 

For the convenience of calculations, we follow the notations introduced in \cite{SSF96}: $\bar{A}(\bar{B})=\mathcal{D}^{-1/2}\mathcal{A(B)}\mathcal{D}^{-1/2}$ and $\bar{X}(\bar{Y})=\mathcal{D}^{1/2}X(Y)$. With this convention, the QRPA equation now has the form:
\begin{eqnarray}
\left(\begin{array}{cc}
\bar{A}&\bar{B}\\-\bar{B}^{*}&-\bar{A}^{*}
\end{array}\right)
\left(\begin{array}{c}
\bar{X}\\\bar{Y}
\end{array}\right)=\omega
\left(\begin{array}{c}
\bar{X}\\\bar{Y}
\end{array}\right)
\label{rqrpa}
\end{eqnarray}
The new form of the QRPA equations are the same as the old ones with X and Y now being replaced by $\bar{X}$ and $\bar{Y}$, so the same technique of solving the QRPA equations could be applied here. The $\mathcal{D}$'s could be obtained approximately with the boson mapping methods introduced in \cite{CDS94}:
\begin{eqnarray}
B^\dagger_{pp'}\rightarrow \sum_{n}B^\dagger_{pn}B_{p'n} \nonumber \\
B^\dagger_{nn'}\rightarrow \sum_{p}B^\dagger_{pn}B_{pn'} \nonumber 
\end{eqnarray}
Where B is defined as $B^\dagger_{\tau\tau'}=\alpha_\tau^\dagger\alpha_{\tau'}$. There are exact derivation of $\mathcal{D}$ with a solvable model\cite{EPS97} which implies that our treatment could have deviations with large particle-particle interaction strength from the exact solutions, this will be discussed later. With this boson mapping approach, we would get the expression for $\mathcal{D}$ as:
\begin{eqnarray}
\mathcal{D}_{pnp'n'}&=&\delta_{pp'}\delta_{nn'}(1 - \sum_{p''}D_{p''np''n'}\sum_{m}\bar{Y}^{m}_{p''n}\bar{Y}^{m}_{p''n'} \nonumber \\
&-& \sum_{n''}D_{pn''p'n''}\sum_{m}\bar{Y}^{m}_{pn''}\bar{Y}^{m}_{p'n''})
\label{Dcoef}
\end{eqnarray}
Eq.\eqref{Dcoef} and eq.\eqref{rqrpa} could be solved iteratively to give us final solutions of the renormalized QRPA(rQRPA) equations.

The expression of $\beta$-decay matrix elements is now a bit different, for the allowed $\beta$-decay matrix elements, we have in the intrinsic system the expression:
\begin{eqnarray}
M_{GT}(E_i)&=&\langle K,i |\tau^-\sigma^{1}_{K}|0^+\rangle \nonumber\\
&=&\sum_{pn} \langle p|\sigma_K |n \rangle \mathcal{D}_{pn}^{1/2} (u_p v_n \bar{X}^i_{pn}+v_p u_n \bar{Y}^i_{pn})
\end{eqnarray} 
Here $K=0,\pm 1$ and $E_i=\omega_{i}^{K}-\omega_{lst}$ follows the definition in \cite{FBS13} with $\omega_{lst}$ the lowest eigenvalues from QRPA equations of all $K^\pi$. The single particle matrix elements $\langle p |\sigma|n\rangle$ are expressed in \cite{YRF08,FBS13}. The First forbidden(FF) matrix elements are in similar style but with much more complicated expressions\cite{FBS13}. The details of the calculations of half-lives in the deformed systems with these calculated matrix elements are also presented in \cite{FBS13}, we will not give the details here.

\section{parameters}

\begin{figure}
\includegraphics[scale=0.35]{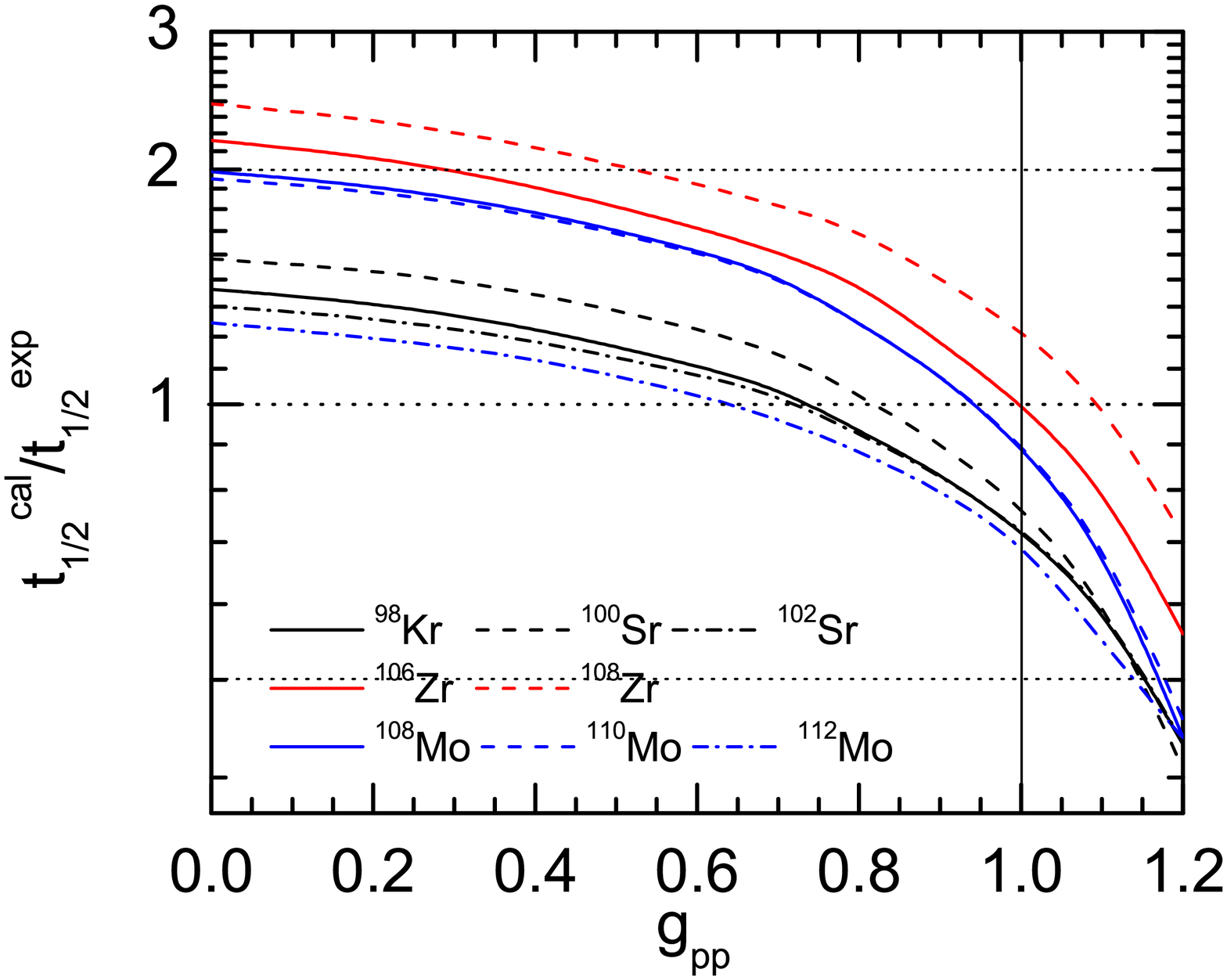}
\caption{(Color online) Illustration of half-life dependence on particle-particle residual interaction strength $g_{pp}$ from rQRPA calculation with realistic forces.}
\label{tpp}
\end{figure}

For the meanfield, we use the same wave functions as in \cite{FBS13} obtained by solving Schr\"odinger equation with Woods-Saxon potentials parametrized as in \cite{FBS13,NBF86}. For the sake of comparison in region II with other results, in this work, for the choice of deformation parameters, if they are available from the measured data\cite{NNDC}, we use the experimental values, otherwise we adopt the predicted values from FRDM model\cite{MNM93} instead of those from HFB17\cite{GCP09} used in \cite{FBS13}. The same strategy is used for the parametrization of nuclear masses-- we use the values from FRDM model\cite{MNM93} unless they are experimentally available\cite{NNDC}, with these masses, we could then get the $\beta$-decay Q-values (the errors from measurements will be considered as we will show later) as well as pairing gaps from the five-point formula\cite{ABB03}. By adjusting the pairing strength parameters $d_{pp}$ and $d_{nn}$ which are overall renormalization factors for the realistic G-matrix elements, we fit these calculated pairing gaps. These choices of parameters deviated from those in \cite{FBS13}, this will also lead to deviation of the final decay rates from those in \cite{FBS13} as shown in fig.\ref{hf}. 

For the residual interaction, as in\cite{FBS13}, we use the G-matrix with realistic CD-Bonn potential. The two most important parameters in our approach is the renormalization strength of residual interactions in particle-hole channel($g_{ph}$) and particle-particle channel($g_{pp}$). The former is related to the position of GTR and doesn't affect the low-lying structure, for this parameter, so we use the value as explained in \cite{FBS13}(This is also partially due to the fact that rQBA doesn't affect particle-hole interactions). The latter parameter $g_{pp}$ is crucial for $\beta$-decay half-lives, we present our strategy for this parameter in detail later.  Before discussing this, we first concentrate on another parameter $g_A$ in nuclear system. As it is known that the measured GT strength amounts only about $60\%$ to the model-independent Ikeda sumrule from various experiments, this raises a problem about the quenching of the axial vector constant $g_A$, we will not go into details about the long bothering question about origin of this quenching, but simply use a value $g_A=0.75g_{A0}$ extracted from various experiments such as\cite{Gue11}. 
%added part 
It is known that rQRPA would violate the Ikeda sumrule\cite{SSF96}, but our calculation shows that this suppression of GT strength affects more of the resonance than the low-lying part. Hence for $\beta$-decay which is related to the low-lying part of GT strength, a quenching factor the same of that for QRPA could be used. 

In this work, as we have mentioned above, we focus on two deformed regions which are on the r-process path, for Kr-Mo region(Region I), $\beta$-decay half-lives of most nuclei have been measured recently \cite{Nis11,Lor15}, while for the important rare earth region (region II), at present, most neutron-rich nuclei are still out of the reach of the experimental facilities, nevertheless there are urgent needs for accurate predictions of their weak decay properties as these are important inputs for r-process abundance simulations with extra importance on the formation of the so-called rare earth peak\cite{MMS11,MMS12}. In this work we adopt model space larger than that used \cite{FBS13}(N=0-6 {\it vs.} N=0-5) for region I and for region II, a even-larger model space of N=0-7 is used to eliminate possible errors from model space truncation. Our calculation show that, enlargement of model space in region I causes just a small change about several percents for decay rates, this is small relative to the errors of the many-body model--QRPA itself. The reason is that most low-lying GT or FF transitions occurs between levels near Fermi surface, contributions from levels far below or far above Fermi energy are relatively small especially for low-lying states. In \cite{FBS13}, we have found sensitive dependence of the half-lives on the particle-particle strength $g_{pp}$, this calls for accurate determination of this parameter as mentioned above. In this work, we follow the treatment in \cite{FBS13} by finding proper value of $g_{pp}$ from the dependence of $t_{1/2}$ on $g_{pp}$, as is illustrated in fig.\ref{tpp}. If we compare the current $t^{theo.}_{1/2}/t^{exp.}_{1/2}-g_{pp}$ curves with those in fig.2 of \cite{FBS13} we will find that the curves have been stretched out at $x$ direction, this is caused by the relaxation of over-correlation of residual interactions in particle-particle channel with the inclusion of PEP. The values of $g_{pp}$ in  QRPA calculations are possibly affected be several factors such as the model space or PEP {\it etc.}. In this work, the model space issue has been carefully handled by using a pretty large model space. With this treatment, our arguments are now that the major reason of the necessity of $g_{pp}$ in QRPA calculations is than the over-correlation from PEP violation. So in this sense, when PEP effects have been considered, we could use bare G-matrix elements of realistic forces in QRPA calculations, and the main errors produced by this treatment are from the approximation of our boson mapping method instead of exact realization of PEP. The advantage of this treatment is now straightforward, that we could handle the regions where fitting of the parameter $g_{pp}$ are nearly impossible due to the lack of experimental data. To see the validity of above arguments, we could check the errors in fig\ref{tpp} when the bare G-matrix is used in particle-particle channel, namely, $g_{pp}=1$. Following the vertical lines in the figure, we see that the calculated half-lives deviate the experimental ones by at most 30 percents and the general trend is that the theory slightly over-predicted the decay rates, the reason is that residual interaction in particle-particle channel is still a bit over-correlated with boson mapping approximation from conclusions in \cite{EPS97} where they have compared the exact treatment and boson mapping technique with a solvable model. With this in mind, we could draw a conclusion that violation of PEP is the main source of a heavy renormalization of residual interaction in previous QRPA calculation. We will check the validity of this conclusion further with more nuclei in both regions and give predictions of the weak decay properties for more even-even isotopes in the next section.

\section{results and discussion}

\begin{figure*}
\includegraphics[scale=0.32]{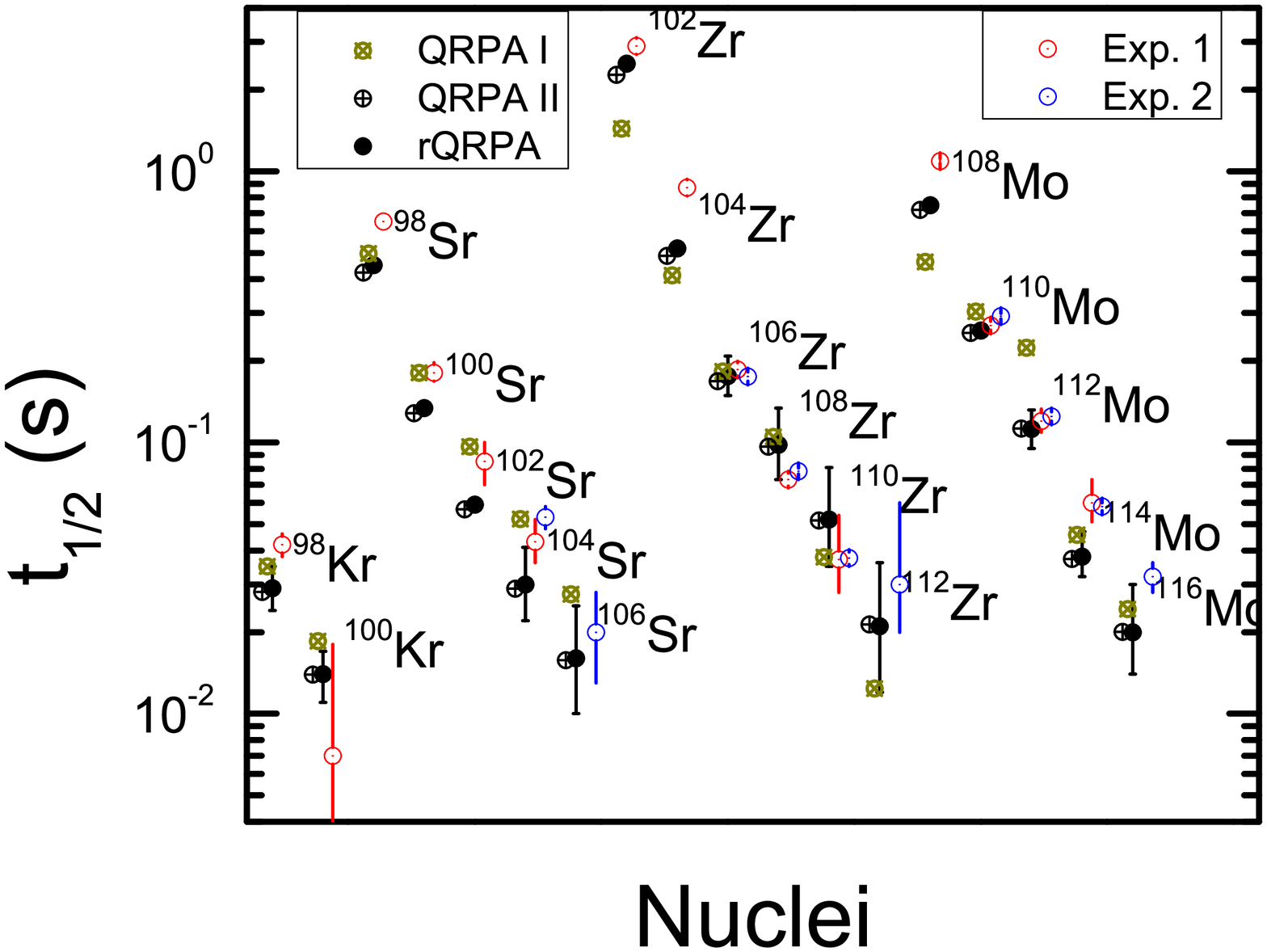}
\includegraphics[scale=0.32]{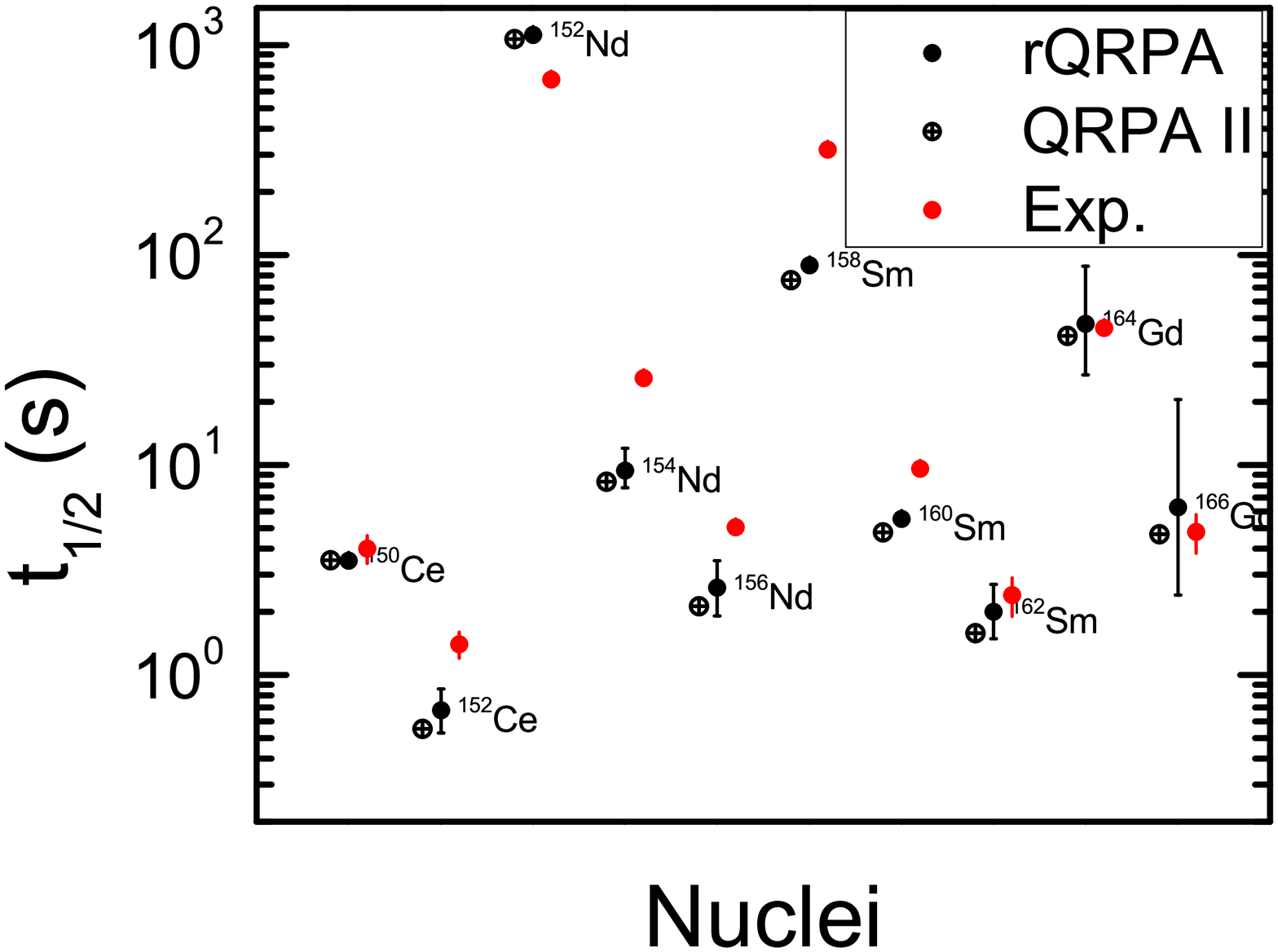}
\caption{(Color online) Comparison of results among calculated and measured half-lives of even-even isotopes in Kr-Mo region (left panel) and rare earth region (right panel). For left panel, QRPA I are results from \cite{FBS13} with QRPA calculations and 'QRPA II' are QRPA calculations with new mass and deformation parameters used in this work. 'rQRPA' are results with PEP in this work. "Exp1" are measured rates from \cite{Nis11} or NNDC if not presented and \cite{NNDC}, "Exp 2" are from recent measurements in \cite{Lor15}. For the right panel, the "rQRPA" are from this work and "Exp." from NNDC\cite{NNDC}. The error bars of our theoretical calculations come from error bars of Q-values from NNDC\cite{NNDC}.}
\label{hf}
\end{figure*}

As we have shown above, the renormalization parameter $g_{pp}$ introduced in previous QRPA calculations are somehow due to the violation of PEP with large enough model space. This would lead to the conclusion that for rQRPA with a large enough model space, using a bare G-matrix in particle-particle channel could be an optimal choice. To see the robustness of this conclusion, we make comparisons between theoretical calculations and experimental results for more isotopes. First we concentrate on region I, where there have been recent new evaluated data\cite{Nis11} as well as old ones\cite{Lor15}. The new data improves the accuracy for many isotopes in this region, for most nuclei the measured half-lives keep the same, the deviations of the new data to the old one are small. At the meantime, the new Ref.\cite{Lor15} also presents us more decay rates for nuclei neutron-richer. The results of QRPA calculations taken from\cite{FBS13} are compared to QRPA calculation with the same formalism but with different parameters of masses and deformations, as well as the improved results from rQRPA, this arrangement is to separate the changes due to PEP from those caused by modifications of parameters. For rQRPA calculations, we also introduce the errors of Q-values with newest evaluations from NNDC-database\cite{NNDC}, the Q-values change a bit since \cite{FBS13} where the measured Q-values are taken from Mass-evaluation 2003\cite{ABB03}, this causes deviations for some nuclei. For masses those are experimentally unavailable, we use values from Ref.\cite{MNM93} as mentioned above. The new parameters of masses and deformations cause changes to our final results, especially the mass parameters, as it is related to the pairing parameters $\Delta$ and $\beta$-decay Q values, the effect of another parameter-- the deformation on QRPA calculations is discussed in \cite{SP10}. Deviations of about a factor of two for theories are encountered for some nuclei, but for those less neutron-rich, the rates basically keep the same, this can easily be interpreted from less modification of mass parameters for these nuclei. A well fitted $g_{pp}=0.75$ values in \cite{FBS13} basically reproduce the experimental results for even-even nuclei as we see from fig.1 of \cite{FBS13}. The comparison between QRPA and rQRPA results in this work with the same set of parameters consolidates our conclusion in previous section that a bare G-matrix elements used in rQRPA are suitable, provided the fact that results using this arrangement agree with QRPA results with previously fitted $g_{pp}$ nearly exactly. This on the other hand tells us that renormalization parameters $g_{pp}$ previously used for realistic forces comes mainly from the negligence of PEP in the calculation.

Comparing the calculated results with the measurements, we find that for most nuclei the agreement is satisfying except a few Sr isotopes. The errors of the calculation are generally with in a factor of two and for most isotopes the deviations are even smaller. The uncertainties from current mass measurements produce large errors for the final rates, improvement of mass evaluations or empirical mass assessments are needed to enhance the prediction power of current $\beta$-decay theory. Inclusion of more isotopes generally doesn't change our conclusion draw for less isotopes in fig.\ref{tpp}, at this region, the decay half-lives are from second to milliseconds, the general errors follows those in fig.3 of \cite{FBS13}. This error is important as most nuclei along the r-process path are with half-lives of this magnitude. And it is reasonable to estimate that the general error of QRPA calculations for these r-path nuclei should be of the similar magnitude. This estimation would help determine the uncertainty analysis in r-process simulations such as those done by Ref.\cite{MSF15}.

\begin{figure*}
\includegraphics[scale=0.65]{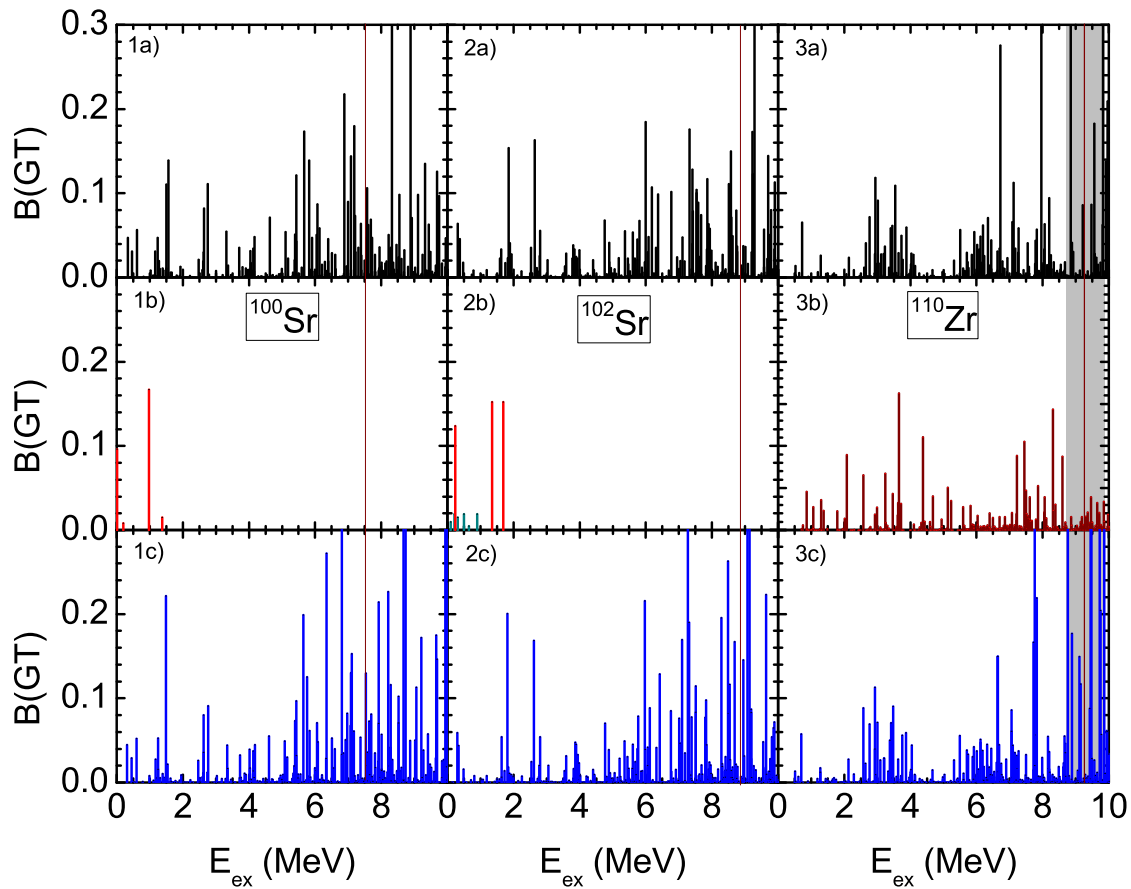}
\caption{(Color online) The B(GT) strength distributions for three nuclei $^{100}$Sr, $^{102}$Sr and $^{110}$Zr. The black droplines are results from QRPA calculations(panels marked with a) and blue lines for rQRPA calculations marked with c). 1b) and 2b) are experimental strength extract from \cite{NNDC}, where red lines are GT transitions and green lines are "probably be GT" transitions. 3b) is taken from \cite{FBS13} from QRPA calculations with parameters different from current calculation. The vertical lines are the Q-values, for $^{110}$Zr, the shaded area denotes the uncertainty of the Q-value.}
\label{strdis}
\end{figure*}

To understand how the the inclusion of PEP changes QRPA calculations and the reason of the similarity of QRPA and rQRPA results for decay rates with specifically fitted $g_{pp}$, we need to compare the detailed $\beta$-decay strength distributions. As FF decays are weak in this region\cite{FBS13}, we compare only the GT strength. From the plotted graph fig.\ref{strdis}, one could find that QRPA with renormalized $g_{pp}$ basically reproduce the $\beta$-strength of rQRPA at low energy region for states with excitation energies below 4 MeV, one-to-one correspondence of the low-lying states has been observed, {\it e.g.} the three states below 1MeV with the same transition structure for $^{100}Sr$ or the two states with large $B(GT)$ values near 2MeV for $^{102}$Sr, as well as the transition around 1MeV for $^{110}$Zr. This correspondence explains the well behaved $\beta$-decay properties from QRPA calculations when PEP is absent. At the meantime, when one goes to the higher energy region with excitation energies above 4MeV, the deviation starts to emerge, GT strength has been relocated more at this energy region for Sr isotopes but less for $^{100}$Zr when PEP is included. These deviations are not important for the decay rates of the isotopes we presented, since their phase space factor is too small to affect the total decay width. But if the Q-values become much larger, then they may play a role when their phase space factors become comparable to those of low-lying strength. In this sense, inclusion of PEP may improve the accuracy of decay properties of much more exotic isotopes. For $^{110}$Zr, new and old QRPA calculations are compared, their difference comes from the different deformation from two models as this is the only difference of parametrization in the two calculations, the relocation of strength from changes of deformation have been discussed also in \cite{FBS13} fig.4 or in \cite{SPF03} for Skyrme calculations, together with this work, one could come to the conclusion that accurate prediction of deformation is vitally important for QRPA calculations. For the two Sr isotopes, we have detailed decay data available, this helps us understand the difference between theory and experiment. The experimental strength is extract from their decay scheme, so only low-lying strength can be obtained as the contributions of high-lying states to decay width have been suppressed by their too small phase space. For each experimental strength below excitation energies of 2MeV, we could always find correspondence from QRPA calculations, {\it e.g.} peak around 1MeV for $^{100}$Sr and two huge transitions around 2MeV. Theories generally predict the structure of the GT strength transitions. Deviations between theory and experiments are inevitable as we could observe, since QRPA is a rough approximation to the exact solutions of nuclear many-body problems. From these graphs, one would naturally raise the problem, whether the deviations could be reduced by adjusting the values of $g_{pp}$. We will show the answer is negative, from our calculations, we draw the conclusion that increasing $g_{pp}$ would shift more strength to low-lying states and also $1^+$ states to much lower energy, {\it vice versa}. In this sense, if one increases the interaction strength in $pp$ channel, he could have the $1^+$ energies agree with experiments, but the deviations of strength then become much worse, on the other hand, if we reduce the strength $g_{pp}$, we could have better agreement of strength but worse for excitation energies. So from such analysis, one could see that the deviations between theories and experiments could be related to other issues as single particle energies and wave functions or deformation but less from $g_{pp}$. The possibility of whether the deviation could be reduced by exact treatment of PEP, such as\cite{EPS97}, still needs investigation.

\begin{figure*}
\includegraphics[scale=0.65]{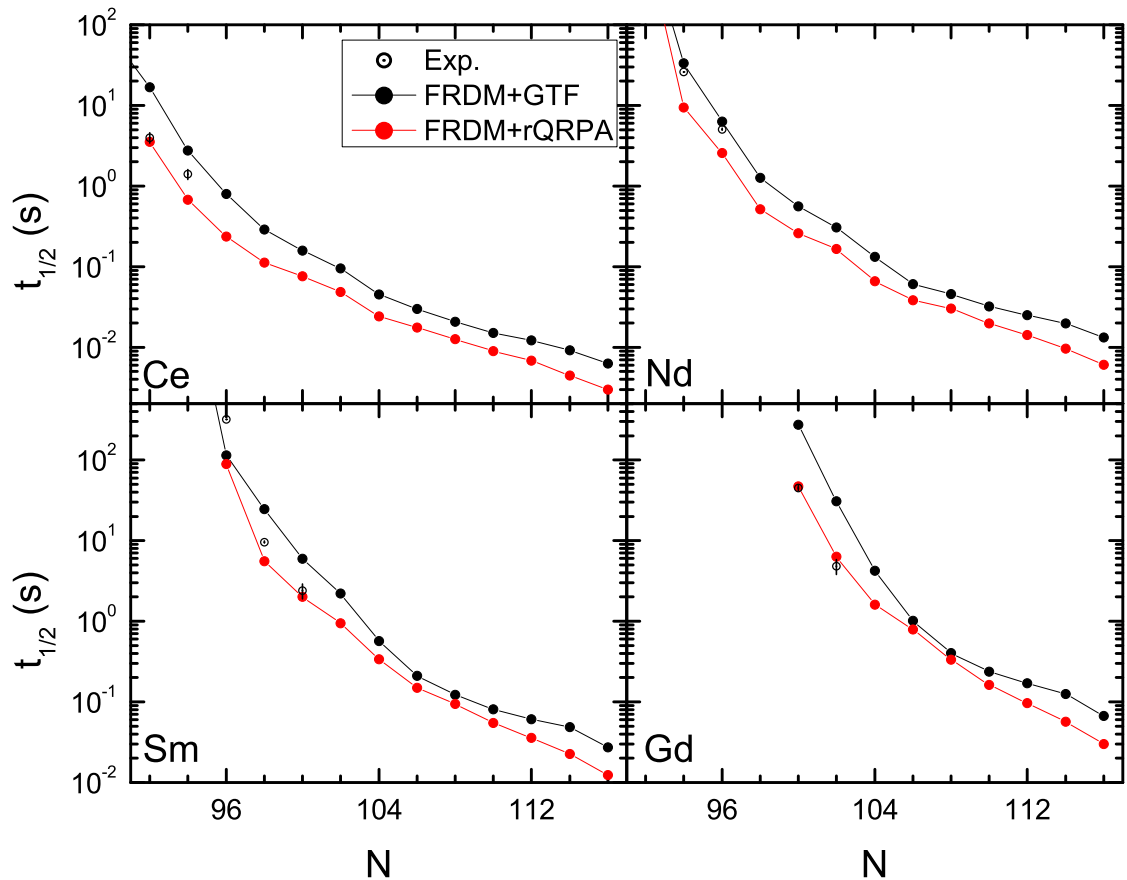}
\caption{(Color online) Prediction of half-lives of even-even nuclei in rare earth regions from rQRPA calculations. Comparison of our results with decay rates from \cite{MPK03} and the experiment results\cite{NNDC} are also presented.}
\label{hfr}
\end{figure*}

We now turn our attention to the less explored region II which is also the target of in-constructing FRIB facilities, see {\it e.g.}\cite{Wre15}. This region is not well explored, neither theoretically nor experimentally, only a few calculations have been done in this region especially for the neutron-rich isotopes, such as results from\cite{MPK03} and from Skyrme calculations\cite{MTZE14} with tensor forces, relatively small errors have been achieved from their calculations. In fig.\ref{hf}, our attempts are also presented. 
% added part
We also compared the rQRPA results with QRPA results, but now for QRPA, different values of $g_{pp}=0.63$ is used. These two sets of results agree with each other as in Region I. These different values of $g_{pp}$ tell us that without rQRPA, the fitting of $g_{pp}$ is needed, this is usually difficult for deformed regions lack of experimental data.
% added part
All the known half-lives of those isotopes in this heavily deformed region are longer than 1 second, these isotopes are with much smaller Q values compared with region I we discussed above. Since QRPA are usually with an error about several hundred keV up to MeV for the prediction of excitation energies, the calculated half-lives could be with less accuracy compared to those of previous region where we could have an error of at most a factor of 2 for all isotopes. For these relatively long-lived isotopes, the deviations may be a bit larger than those in region I but still acceptable. It is difficult to claim that our rates are much more precise than other methods with such limited data. In fig.\ref{hf}, we have comparisons for several isotopes between our calculations and experiments, the difference is not so drastic, a factor of two deviation can be obtained for most nuclei except two long-lived isotopes--$^{154}$Nd and $^{158}$Sm. As compared to FRDM rates from fig.\ref{hfr}, we see that we have better agreements for most Sm and Gd isotopes but they have better predictions for Nd isotopes, and similar errors for Ce isotopes are obtained, our rates are faster and theirs are slower. Also large errors are encountered in our calculation with the uncertainties from mass measurements, this makes comparison of the results less clear and affects an effective estimation over the errors and reliabilities of the theory. The errors could be further reduced with the reduction of uncertainties in nuclear mass data which could be done after FRIB\cite{Wre15}. 

Our new results are served as new alternatives for the nuclear input of r-process simulations. So we did calculations for more nuclei and make comparisons with existing results. The results in\cite{MPK03} have been widely used in astrophysics community and a direct comparison with them could give us some hints over the final simulation. A large deviation of the two calculations has been observed, our rates are much faster than theirs, the reason was explained in \cite{FBS13} that the absence of interactions in particle-particle channel gives overestimation on the half-lives with FRDM model. A factor of 2 differences between theories have been observed for almost all nuclei here. The consequence of these new rates on r-process simulations still needs investigations while rates of all kinds of nuclei can be obtained. This requires us to extend our formalism to odd-A and odd-odd nuclei with improved accuracies, as previous agreement for these odd nuclei from QRPA calculations in\cite{FBS13} is really poor. After these have been done, we could make an estimation on how the new rates affect the abundance pattern and would shed lights on the formation of rare earth peak.  

\section{conclusion}
In this work we have introduced the rQBA into the QRPA calculations with realistic forces. We found that with the new commutation relation, the over-correlation of residual interactions at $pp$ channel is eliminated, hence the need of renormalization of these interaction strength can be neglected. With the new calculation, we come to satisfying agreement between calculation and measurement for both region I and region II for neutron-rich even-even isotopes. We also give predictions of half-lives of much neutron-richer nuclei in rare earth region which are important according to simulations from \cite{MMS11,MMS12}. These calculations would help solve the the long debating problem on the formation of rare earth peak of the solar element abundance.

\begin{acknowledgements}
The author would like to thank Prof. B. A. Brown for helpful discussions and comments. I would like to thank also Prof. J. Engle and Mr. T. Shafer for useful discussions on this topic and hospitality during my visit to UNC-Chapel Hill. Correspondence with Dr. M. Mumpower is also acknowledged. This work is supported by National Natural Science Foundation of China under Grant No. 11505078.
\end{acknowledgements}

\end{document}